\documentclass[showpacs]{revtex4}

\usepackage{graphics}
\usepackage{amssymb}
\usepackage{amsmath}

\usepackage{soul}
\usepackage{color}
\usepackage{graphics}
\usepackage{amsmath,amssymb,amsfonts}
\usepackage{bm}
\usepackage{latexsym}

\begin{document}

\title{Conformal symmetries of the energy-momentum tensor of spherically symmetric static spacetimes}

\author{Ugur Camci}

\affiliation{ Department of Chemistry and Physics, Roger Williams University, One Old Ferry Road, Bristol, Rhode Island \mbox{02809, USA}}

\email{ucamci@rwu.edu, ugurcamci@gmail.com}

\author{Khalid Saifullah}

\affiliation{Department of Mathematics, Quaid-i-Azam University, Islamabad,
Pakistan}

\email{saifullah@qau.edu.pk}

\date{\today}

\begin{abstract} Conformal matter collineations of the energy-momentum tensor for a general spherically symmetric static spacetime are studied. The general form of these collineations is found when the
energy-momentum tensor is non-degenerate, and the maximum number of independent conformal matter collineations is \emph{fifteen}. In the degenerate case of the energy-momentum tensor it is found that these collineations have infinite degrees of freedom. In some subcases of degenerate energy-momentum, the Ricci tensor is non-degenerate, that is, there exist non-degenerate Ricci inheritance collineations.
\end{abstract}



\maketitle

\section{Introduction}
\label{INT}

\noindent Recent observations indicate that the universe contains black holes whose horizons are rotating at a speed close to that of light. General relativity (GR) suggests that the dynamics near the horizon of such black holes is governed by a strong infinite-dimensional conformal symmetry - similar to the one seen near the critical points of different condensed matter systems. Researchers have explored possible observational consequences of such a symmetry \citep{AS}.

Symmetries and conformal symmetries play a very important role in mathematical physics. One of the fundamental symmetries on a Riemannian manifold is that of the metric tensor $\mathbf{g}$
written mathematically as \cite{katzin}
\begin{equation}
\pounds_{\xi}{\bf g} = 2 \sigma {\bf g}. \label{confm}
\end{equation}
Here $\sigma$ is the conformal factor and $\pounds_{\xi}$ represents the Lie derivative operator relative to the vector field ${\xi}$, which gives isometries or Killing vectors (KVs) if the $\sigma$ is zero, homothetic motions (HMs) if it is a constant, and conformal Killing vectors (CKVs) if it is a function of the coordinates $x^a$. In component form we can write the above equation as
\begin{equation}
g_{ab,c} \xi^c + g_{ac} \xi_{,b}^c + g_{cb} \xi^c_{,a} = 2 \sigma g_{ab}. \label{isom1}
\end{equation}
Apart from the metric tensor the other quantities fundamental to the Einstein field equations (EFEs)
\begin{equation}
R_{ab} - \frac{1}{2} R g_{ab} = \kappa T_{ab}, \label{efes}
\end{equation}
are the stress-energy tensor, ${\bf T}$, which describes the matter field in the manifold, the Ricci tensor, $\bf R$, which is a contraction of the curvature tensor, and the Ricci scalar $R$. In Eq. \eqref{efes}  $\kappa$ is the coupling constant defined by $\kappa = 8 \pi G / c^4$, $G$ and $c$ are Newton's gravitational constant and the speed of light.  Thus, the
symmetries of both the stress-energy tensor and the Ricci tensor, play a significant role. These symmetries known as matter collineations (MCs) and Ricci collineations (RCs) \cite{hall-book}, respectively, satisfy
the equations
\begin{equation}
\pounds_{\xi}{\bf T} = 0 , \,\,\,\  \text {and} \,\,\,\
\pounds_{\xi}{\bf R} = 0 . \label{RC}
\end{equation}
Similarly, one can define collineations for the curvature and Weyl tensors \cite{hall-book}.

Solutions of EFEs can be classified by requiring these symmetries and thus a complete list of metrics having certain symmetry can be obtained \cite{ESEFEs}. Spacetimes have been classified
on the basis, for example, of KVs \cite{bq1987,qz1988,qz1995}, HMs \cite{dz1997}, CKVs \cite{maartens1995,keane2004}, RCs \cite{tm1990,bk1993,fqz1995,hall1996,qz1998,ziad2003,saif2003,ugur1,cb2002,ugur2,ugur3,tsamparlis1} and MCs \cite{ugur4,sharif2003,sharif2004,ugur5,ugur6}. This also provides a way to find new solutions of EFEs which are otherwise very difficult to solve. These collineations have been generalized to define what are called \emph{conformal collineations} (or \emph{inheritance collineations} \cite{duggal1,duggal11,duggal2}). Thus we obtain conformal matter collineations (CMCs)
\begin{equation}
\pounds_{\bf X}{\bf T} = 2 \psi(x^a) {\bf T}, \label{cmc}
\end{equation}
or conformal Ricci collineations (CRCs) defined by
\begin{equation}
\pounds_{\bf Y}{\bf R} = 2 \phi(x^a) {\bf R}. \label{CRC}
\end{equation}

Conformal symmetry is physically significant as CKVs, for example, generate constants of motion along the null geodesics for massless particles which are conserved quantities. On the other hand, it is of mathematical interest to obtain classification by conformal collineations and to investigate their relation with collineations. Though there has been a good amount of literature on the study of CKVs, the interest in conformal collineations is relatively recent. The complete classification of spherically symmetric static space-times by their CRCs when the conformal factor $\phi$ is a constant has been carried out in Ref. \cite{bkk2003}. The CRCs with a non-constant $\phi$ have been studied for the Friedmann-Robertson-Walker spacetimes \cite{cb2002}, the general static spherically symmetric spacetimes \cite{crcsph}, the non-static spherically symmetric spacetimes \cite{bokhari2020} and Kantowski-Sachs spacetimes \cite{hkb2018}.  For pp-waves relationship between CRCs and CKVs has been studied in Ref. \cite{keane2004}. Recently, Akhtar et al. \cite{bokhari2018} have classified static plane symmetric space-times according to CRCs. Further, the CRCs for the Einstein-Maxwell field equations in the case of non-null electromagnetic fields have been investigated as well \cite{faridi}.

In this paper we classify spherically symmetric static spacetimes by their CMCs. The equation \eqref{cmc} for CMCs in component form can be written as
\begin{equation} \label{crc2}
T_{ab,c} X^c + T_{ac} X^c_{,b} + T_{cb} X^c_{,a} = 2 \psi T_{ab} ,
\end{equation}
where $\psi$ is the conformal factor which is a function of all the spacetime coordinates $x^a=(x^0, x^1, x^2, x^3)$. In this paper we use the usual component notation in local charts and a partial derivative will be denoted by a comma. Note that the above equation gives MCs if $\psi =0$, thus the classification of spherically symmetric static spacetimes by MCs \cite{sharif2003} becomes a special case of the classification obtained in this paper. We call a CMC as \emph{proper} if it is neither a KV nor an MC. The set of all CMCs on the manifold is a vector space, but it may be infinite dimensional. If $T_{ab}$ is non-degenerate, i.e. $det(T_{ab}) \neq 0$, then the Lie algebra of CMCs is finite dimensional but if $T_{ab}$ is degenerate, it may be infinite. Thus, in the case of a non-degenerate energy-momentum tensor, i.e. $det(T_{ab}) \neq 0$, we use the standard results on conformal symmetries to obtain the maximal dimensions of the algebra of CMCs as 15. Since $T_{ab}$ describes the distribution and motion of matter contents of a manifold, and mathematically it is very similar to the the Ricci tensor, the study of MCs and CMCs has a natural geometrical as well as physical significance.

In the next section, we setup the CMC equations for static spherically symmetric spacetimes.
In Section \ref{degenerate} these equations are solved when the energy-momentum tensor is degenerate, while in Section \ref{non-degenerate} we obtain results when the tensor is non-degenerate. We find that the degenerate case always gives infinite dimensional Lie algebras of CMCs. We conclude with a brief summary and discussion in Section \ref{conc}. Throughout the paper, we will consider four-dimensional space-times, and space-time indices will be denoted by small Latin letters (e.g., $a$, $b$, $c$,...) and the metric has signature $(+,-,-,-)$.

\section{Equations for Conformal Matter Collineations }
\label{rics}

\noindent We consider a general spherically symmetric static spacetime in the usual spherical coordinates
\begin{equation}\label{metric}
ds^2 =  e^{\nu (r)} dt^2 - e^{\lambda (r)} dr^2 - r^2 \left( d\theta^2 + \sin^2 \theta d\phi^2 \right) .
\end{equation}
The non-vanishing components of the Ricci tensor $R_{ab} = R^c_{\, a c b}$ for this metric are given by
\begin{eqnarray}
& & R_{00} \equiv R_0 (r) = \frac{1}{4} e^{\nu - \lambda} \left( 2
\nu'' + {\nu'}^2 - \nu' \lambda' + \frac{4}{r} \nu' \right),
\label{r0} \\& & R_{11} \equiv R_1 (r) = -\frac{1}{4} \left( 2
\nu'' + {\nu'}^2 - \nu' \lambda' - \frac{4}{r} \lambda' \right),
\label{r1} \\& & R_{22} \equiv R_2 (r) =  \frac{1}{2} e^{-\lambda}
\left[ r (\lambda' - \nu') - 2 \right] + 1, \label{r2}\\& & R_{33}
= \sin^2 \theta  R_{2}, \label{r2-r3}
\end{eqnarray}
and the Ricci scalar $R = R^a_{\,\,a}$ is
\begin{equation}\label{ricciscalar}
R = \frac{e^{-\lambda}}{2} \left[ 2\nu'' + \nu'^2 -\nu' \lambda'
+ \frac{4}{r} (\nu' - \lambda') + \frac{4}{r^2} (1-e^{\lambda})
\right],
\end{equation}
where the prime represents derivative with respect to the radial coordinate $r$. Thus we can write the \emph{Ricci tensor form} as
\begin{equation}\label{ricci-metric}
ds^2_{Ric} \equiv R_{ab} dx^a dx^b =  R_0 (r) dt^2 + R_1 (r) dr^2
+ R_2 (r) \left( d\theta^2 + \sin^2 \theta d\phi^2 \right).
\end{equation}

The metric \eqref{metric} has time-independent coefficients, and using the field equations \eqref{efes} with $c=1$ and $G=\frac{1}{8\pi}$, i.e., $\kappa =1$, the components of the energy-momentum tensor $T_{ab}$ become
\begin{eqnarray}
& & T_{00} \equiv T_0 (r) = \frac{1}{r^2} e^{\nu - \lambda} \left( r \lambda' + e^{\lambda} - 1 \right), \label{t0} \\& & T_{11} \equiv T_1 (r) = \frac{1}{r^2} \left(r \nu'- e^{\lambda} +1 \right), \label{t1} \\ & & T_{22} \equiv T_2 (r) =  \frac{r^2}{4} e^{-\lambda} \left[ 2 \nu''+ {\nu'}^2- \nu' \lambda' + \frac{2}{r} (\nu' - \lambda') \right] , \label{t2}\\& & T_{33} \equiv T_3 (r) = \sin^2 \theta T_{2}. \label{t2-t3}
\end{eqnarray}
Similarly, the matter tensor form can be written as
\begin{equation}\label{matter-metric}
ds^2_{Matter} \equiv T_{ab} dx^a dx^b =  T_0 (r) dt^2 + T_1 (r)
dr^2 + T_2 (r) \left( d\theta^2 + \sin^2 \theta d\phi^2 \right).
\end{equation}
Using the above energy-momentum tensor components, the Ricci curvature scalar given in \eqref{ricciscalar} can be cast into the form
\begin{equation}\label{ricciscalar-2}
R = -e^{-\nu} T_0 + e^{-\lambda} T_1 + \frac{2}{r^2} T_2 \, .
\end{equation}
Further, when we state the energy-momentum tensor components given above in terms of the Ricci tensor components \eqref{r0}-\eqref{r2} we find that
\begin{eqnarray}
& & T_{0} = \frac{1}{2} \left( R_0 +  e^{\nu -\lambda} \, R_1 \right) + \frac{e^{\nu}}{r^2} R_2, \label{t0-n} \\& & T_1 = \frac{1}{2} \left( e^{\lambda-\nu} R_0  + R_1 \right) - \frac{e^{\lambda}}{r^2} R_2, \label{t1-n} \\ & & T_{2} =  \frac{r^2}{2}  \left( e^{-\nu} R_0 - e^{-\lambda} R_1 \right). \label{t2-n}
\end{eqnarray}

In GR, physical fields are described by the symmetric tensor $T_{ab}$ which is the energy-momentum tensor of the field. We have the covariant decomposition/identity for $T_{ab}$ as follows \cite{tp2019}:
\begin{equation}
  T_{ab} = \rho u_a u_b - p h_{ab} + 2 q_{(a} u_{b)} + \pi_{ab} \, , \label{anisot-fluid}
\end{equation}
where $h_{ab} = g_{ab} -  u_a u_b$ is the projection tensor, and the quantities $\rho, p, q_a$ and $\pi_{ab}$ are the physical variables representing the mass density, the isotropic pressure, the heat flux and the traceless stress tensor, respectively, as measured by the observers $u^a$. In the above decomposition, $T_{ab}$ is described by two scalar fields ($\rho,p$), one spacelike vector ($q^a, \, q_a u^a = 0$), and a traceless symmetric 2-tensor ($\pi_{ab}, \, g_{ab} \pi^{ab} = 0$). The irreducible parts of $T_{ab}$ are defined as
\begin{eqnarray}
& &  \rho = u^a u^b T_{ab} \, , \label{rho-emt}  \\ & & p = -\frac{1}{3} h^{ab} T_{ab} \, , \label{p-emt}  \\ & &  q^a = h^{ab} T_{bc} u^c \, , \label{q-emt} \\ & & \pi_{ab} = \left( h_{a}^{\, c} h_{b}^{\, d} - \frac{1}{3} h_{ab} h^{cd} \right) T_{cd} \, , \label{pi-emt}
\end{eqnarray}
where $u^a$ is a timelike unit four vector field  normalized by $u^a u_a = 1$.
The energy-momentum tensor $T_{ab}$ given in \eqref{anisot-fluid} represents the general anisotropic fluid, and reduces to an anisotropic fluid without heat flux if $q^a =0$, an isotropic non-perfect fluid if $\pi^{ab} = 0$, a perfect fluid if $q^a = 0$ and $\pi^{ab} = 0$, and a dust if $p=0, q^a = 0$ and $\pi^{ab} = 0$. GR is a classical theory of relativity, however, in the field equations \eqref{efes}, the classical spacetime geometry is also related to the stress-energy tensor of quantum matter. To overcome this inconsistency, we need to embed GR (or its generalizations) within some quantum mechanical framework, i.e., quantum gravity. For the metric ansatz \eqref{metric}, it is customary to choose the fluid to be at rest because the spacetime is static, i.e., $u^a = u^0 \delta^a_0$. Then, using the normalization condition of the four-velocity, that is, $u^a u_a = 1$, one can find $u^0 = e^{-\nu /2}$. Thus, we find from \eqref{q-emt} that for this choice of observers the heat flux vanishes ($q^a = 0$),  which is expected from the symmetries of the metric. Also, under the latter choice of observers, the remaining physical variables $\rho, p$ and $\pi_{ab}$ that follow from \eqref{rho-emt}, \eqref{p-emt} and \eqref{pi-emt} are
\begin{eqnarray}
& &  \rho = e^{-\nu} T_0 \, , \qquad p = \frac{1}{3} \left( e^{- \lambda} T_1 + \frac{2}{r^2} T_2 \right) \, , \label{rho-p-emt} \\ & & \pi_{00} = 0, \quad  \pi_{11} = \frac{2}{3} \left( T_1 - \frac{e^{\lambda}}{r^2} T_2 \right)  \, , \quad   \pi_{22} = -\frac{r^2}{2} e^{-\lambda} \pi_{11}  \, , \quad  \pi_{33} =  \sin^2 \theta \, \pi_{22} \, . \label{pi-123-emt}
\end{eqnarray}
If the choice of observers as a timelike four-vector field $u^a = u^0 \delta^a_0$ is not appropriate for any reason, then we must apply the normalization condition of the four-velocity with $u^a u_a = -1$ by choosing the four-velocity as $u^a = u^1 \delta^a_1$ which is a spacelike four-vector, since one can always normalize the four-velocity to $\pm 1$. Then we find for the metric \eqref{metric} that $u^1 = e^{-\lambda/2}$. For the spacelike four-vector, the projection tensor has the form $h_{ab} = g_{ab} +  u_a u_b$. Then this choice effects the mass density, the isotropic pressure $p$ and the traceless stress tensor $\pi_{a b}$ due to Eqs. \eqref{rho-emt}, \eqref{p-emt} and \eqref{pi-emt} such that
\begin{eqnarray}
& &  \rho = e^{-\lambda} T_1 \, , \qquad p = - \frac{1}{3} \left( e^{- \nu} T_0 - \frac{2}{r^2} T_2 \right) \, , \label{rho-p-emt-2} \\ & & \pi_{11} = 0, \quad \pi_{00} = \frac{2}{3} \left( T_0 + \frac{e^{\nu}}{r^2} T_2 \right)  \, , \quad   \pi_{22} = \frac{r^2}{2} e^{-\nu} \pi_{00}  \, , \quad  \pi_{33} =  \sin^2 \theta \, \pi_{22} \, . \label{pi-123-emt-2}
\end{eqnarray}

In GR, it is conventional to restrict the possible energy-momentum tensors by imposing energy conditions. The energy conditions for the energy-momentum tensor $T_{ab}$ to represent some known matter fields are the conditions that are coordinate-independent restrictions on $T_{ab}$. In literature, there are five categories of the energy conditions. These are the trace energy conditions (TEC), null energy conditions (NEC), weak energy conditions (WEC), strong energy conditions (SEC) and dominant energy conditions (DEC). The TEC means that the trace of the energy-momentum tensor $T = g^{ab} T_{ab}$ should always be positive (or negative depending on the signature of metric). The NEC mathematically states that $T_{ab} k^a k^b \geq 0$ for any null vector $k^a$, i.e., $k^a k_a =$. On the other hand, the WEC requires that $T_{ab} t^a t^b \geq 0$ for all timelike vectors $t^a$.  The SEC states that $T_{ab} t^a t^b \geq \frac{1}{2} T t^c t_c $ for all timelike vectors $t^a$. The DEC includes the WEC, as well as  the additional requirement that $T^{ab} t_b$ is a non-spacelike vector, i.e., $T_{ab} T^b_{\,\, c} t^a t^c \leq 0$ \cite{carrol,capo2015}.
For a perfect fluid, the energy conditions are described as
\begin{eqnarray}
& & {\rm TEC:} \quad \rho - 3 p \geq 0 \, ; \qquad {\rm NEC:} \quad \rho + p \geq 0 \, ; \qquad {\rm WEC:} \quad \rho \geq 0 \, , \,\, \rho + p \geq 0 \, ; \\
& &  {\rm SEC:} \quad \rho + 3 p \geq 0 \, , \,\, \rho + p \geq 0 ; \quad {\rm DEC:} \quad \rho \geq 0 \, , \,\, \mid p \mid \leq \rho \, .
\end{eqnarray}
Thus it is seen that the energy conditions are simple constraints on various linear combinations of the energy density $\rho$ and the pressure $p$. The matter including both positive energy density and positive pressure, which is called "normal" matter, satisfies {\it all} the standard energy conditions. On the contrary, "exotic" matter violates any one of the energy conditions. For example, the SEC is satisfied by "all known forms of energy", but not by the dark energy where $p = -\rho$. We note that for the static and spherically symmetric metric \eqref{metric} one can find the null vector as $k^a =  e^{-\nu /2} \delta^a_0 + e^{-\lambda /2} \delta^a_1$ by using the condition $k^a k_a = 0$.
Then the NEC for the general anisotropic fluid \eqref{anisot-fluid} becomes
\begin{equation}
  \rho + 3 p - \frac{2}{r^2} T_2 \geq 0 \, ,  \label{nec-anf}
\end{equation}
which yields the perfect fluid energy condition taking $T_2 = p r^2$.

For the perfect fluid, it is easily seen that $T_0 = \rho e^{\nu}, \,\, T_1 = p e^{\lambda}, \,\, T_2 = p r^2$, i.e., $T_2 = r^2 e^{-\lambda} T_1$, and $T_3 = sin^2 \theta T_2$, which yields $\pi_{ab} = 0$ as it should be, and these give rise to the following relations
\begin{equation}
R_{0} = \frac{e^{\nu}}{2} (\rho + 3 p), \quad R_{1} =
\frac{e^{\lambda}}{2} ( \rho - p), \quad R_{2} = \frac{r^{2}}{2}( \rho - p ), \label{r012}
\end{equation}
where $\rho$ and $p$ are, respectively, density and pressure of the fluid, which are
\begin{equation}
\rho =  \frac{e^{-\lambda}}{r^2} \left( r \lambda' + e^{\lambda}
-1 \right), \qquad p =  \frac{e^{-\lambda}}{r^2} \left( r \nu' -
e^{\lambda} +1 \right). \label{rho-p}
\end{equation}
Further, the condition of isotropy of the pressure for the perfect fluid matter yields that
\begin{equation}\label{iso-p}
2 \nu'' +\nu'^2 -\nu' \lambda' - \frac{2}{r} \left(\nu' + \lambda'
\right) + \frac{4}{r^2} \left( e^{\lambda} -1 \right) = 0,
\end{equation}
which is equivalent with
\begin{equation}
r^2 R_1 - e^{\lambda} R_2  = 0 \, .
\end{equation}
In this case the energy conditions for a barotropic equation of state $p = p(\rho)$ are given by
\begin{equation}
\rho >0, \qquad 0 \leq p \leq \rho, \qquad 0 \leq \frac{dp}{d\rho} \leq 1 \, .
\end{equation}
The linear form of a barotropic equation of state is given by $p = w \rho$ where $w$ is the equation of state parameter.  For $w =0, 1/3$ and $1$, we get \emph{dust, incoherent radiation} and \emph{stiff matter}, respectively.

For the spherically symmetric static spacetimes \eqref{metric}, Eq. \eqref{crc2} takes the form:
\begin{eqnarray}
& & T'_{i} X^1 + 2 T_{i} X^i_{,i}  = 2 T_{i} \psi, \,\,\ i=0,1,2 \label{crc-a}
\\& & T_{2} \left( X^2_{,2} -\cot\theta X^2 - X^3_{,3} \right)= 0, \label{crc-d}
\\& & T_0 X^0_{,j} + T_j X^{j}_{,0} = 0, \,\,\,\ j=1,2,3\label{crc-e}
\\ & & T_j X^{j}_{,k} + T_{k} X^{k}_{,j} = 0, \,\,\, j,k = 1,2,3 \,\,\, (j \neq k) \label{crc-f}
\end{eqnarray}
In the above equations the summation convention is not assumed. For the non-degenerate energy-momentum tensor $T_{ab}$, after some tedious calculations similar to those done in Ref. \cite{crcsph}
we see that the general solution of equations \eqref{crc-a}-\eqref{crc-f} can be written as
\begin{eqnarray}
X^0 &=& \frac{T_2}{T_0} \left[ \sin\theta \left( A'_1 \sin\phi - A'_2 \cos\phi \right) + A'_3 \cos\theta \right] + A_4(t,r), \label{x0} \\ X^1 &=& \frac{T_2}{T_1} \left[ \sin\theta \left( A'_1 \sin\phi - A'_2 \cos\phi \right) + A'_3 \cos\theta \right] + A_5(t,r), \label{x1} \\ X^2 &=& - \cos\theta \left[ A_1 \sin\phi - A_2 \cos\phi \right]+ A_3 \sin\theta + a_1 \sin\phi - a_2 \cos\phi, \label{x2} \\ X^3 &=& -csc\theta \left[ A_1 \cos\phi + A_2 \sin\phi \right]+ \cot\theta (a_1 \cos\phi + a_2 \sin\phi) + a_3, \label{x3}
\end{eqnarray}
with the conformal function given by
\begin{eqnarray}
& & \psi = \left( \frac{T'_{2}}{2 T_{2}} A'_1 + A_1 \right) \sin\theta \sin\phi - \left( \frac{T'_{2}}{2 T_{2}} A'_2 + A_2 \right) \sin\theta \cos\phi  \nonumber \\ & & \qquad + \left( \frac{T'_{2}}{2 T_{2}} A'_3 + A_3 \right) \cos\theta + \frac{T'_{2}}{2 T_{2}} A_5 (t,r), \label{psi}
\end{eqnarray}
where $A_{\ell} (t,r), \,(\ell = 1,2,3,4,5)$, are integration functions and $a_j \,(j=1,2,3)$ are constant parameters, which give the \emph{three} KVs of spherically symmetric spacetimes
\begin{eqnarray}
& & \mathbf{X_1}=\sin\phi \, \partial _{\theta } + \cos\phi \cot\theta \, \partial _\phi ,  \quad
\mathbf{X_2}= \cos\phi \, \partial _\theta - \sin\phi \cot\theta \, \partial _\phi ,  \quad \mathbf{X_3}=\partial _{\phi} .
\label{sphmin}
\end{eqnarray}
Further, the functions $A_{\ell} (t,r) $ in the above equations \eqref{x0}-\eqref{psi} are subject to the following constraint equations
\begin{eqnarray}
& & \left( \sqrt{\frac{T_{1}}{T_{2}}} A_5 \right)' = 0, \label{cnst-a}
\\& & T_{0} A_4' + T_{1} \dot{A_5} = 0, \label{cnst-b} \\& & 2 \dot{A_4} + \left( \frac{T'_{0}}{T_{0}} - \frac{T'_{2}}{T_{2}} \right) A_5 = 0, \label{cnst-c} \\& & \left( \sqrt{\frac{T_{2}}{T_{0}}}
\dot{A}_j \right)' = 0, \label{cnst-d}\\& & \ddot{A}_j + \frac{T_{0}}{2 T_{1}} \left( \frac{T'_{0}}{T_{0}} - \frac{T'_{2}}{T_{2}} \right) A'_j - \frac{T_{0}}{T_{2}} A_j = 0,  \label{cnst-e}\\& & A''_j + \frac{1}{2} \left( \frac{T'_{2}}{T_{2}} - \frac{T'_{1}}{T_{1}} \right) A'_j - \frac{T_{1}}{T_{2}} A_j = 0,  \label{cnst-f}
\end{eqnarray}
where the dot represents the derivative with respect to time $t$, and $j=1,2,3$. When we solve the above constraint equations for possible cases of non-degenerate energy-momentum
tensor $T_{ab}$, we obtain the corresponding CMCs for the spherically symmetric static spacetimes (see Section \ref{non-degenerate}). In the following section we find CMCs for the degenerate energy-momentum tensor of the spherically symmetric static spacetimes.

\section{Conformal Matter Collineations for the Degenerate Matter Tensor}
\label{degenerate}

\noindent If the energy-momentum tensor is degenerate, that is, $det(T_{ab})= 0$, then we have the following \emph{four} possibilities: {\bf (D-A1)} all of the $T_{a}\, (a=0,1,2,3) $  are zero; {\bf (D-A2)}
one of the $T_{a}$ is nonzero; {\bf (D-A3)} two of the $T_{a}$ are nonzero; {\bf (D-A4)} three of the $T_{a}$ are nonzero.

\bigskip

\noindent {\bf Case (D-A1)}. This case corresponds to the vacuum (such as the Schwarzschild) spacetime in which every vector is a CMC.

\bigskip

\noindent {\bf Case (D-A2)}. In this case, we have the subcases such that: {\bf (D-A2-i)} $T_{0} \neq 0, \,\, T_{j} = 0, \,\, (j = 1,2,3)$;\,\, {\bf (D-A2-ii)} $T_{1} \neq 0, \,\, T_{k} = 0, \,\,
(k = 0,2,3)$.


{\bf Subcase (D-A2-i)}. In this subcase we find
\begin{eqnarray}
& X^0 = X^0(t), \quad X^1 = (\psi - \dot{X}^0)\frac{2 T_{0}}{T'_{0}} \, , \quad X^{\alpha} = X^{\alpha} (x^a) \, ,  \label{a2i1}
\end{eqnarray}
where $T'_{0} \neq 0$ and $\alpha = 2,3$. If $T'_{0} = 0$, i.e., $T_{0} =c$ (a constant), then the CMCs takes the following form
\begin{eqnarray}
& X^0 = \int{\psi dt} + a, \quad X^j = X^j (x^a), \label{a2i2}
\end{eqnarray}
where $a$ is an integration constant. The corresponding Lie algebra of the vector fields in this subcase is infinite dimensional because the vector fields given in \eqref{a2i1} and \eqref{a2i2} have arbitrary components. For this subcase, using \eqref{t0-n}-\eqref{t2-n} we have
\begin{equation}
T_0 = -2 R_0, \quad R_1 = e^{\lambda - \nu} R_0, \quad R_2 = r^2
e^{-\nu} R_0,  \label{d-a2i}
\end{equation}
which shows that all Ricci tensor components are nonzero, e.g. the Ricci tensor in this subcase is non-degenerate. Furthermore, in this subcase we have the equation of state $p=0$ (dust) for the perfect fluid.


{\bf Subcase (D-A2-ii)}. Considering the constraints of this subcase, we obtain that
\begin{eqnarray}
& X^1 = \frac{1}{\sqrt{\mid T_{1} \mid}} \left[ \int{\psi\,\sqrt{\mid T_{1} \mid} dr} + a_1 \right], \qquad X^k = X^k (x^a), \label{a2ii}
\end{eqnarray}
where $a_1$ is a constant of integration. Here, we again have infinite dimensional Lie algebra of vector fields. Using \eqref{t0-n}-\eqref{t2-n}, we find
\begin{equation}
R_0 = e^{\nu -\lambda} R_1 \, , \qquad R_2 = - r^2 e^{-\lambda} R_1 \, , \qquad T_1 = 2 R_1 \, ,  \label{d-a2ii}
\end{equation}
which are independent relations for any form of the energy-momentum tensor.
It is obvious that for this case the choice of observers as a timelike four-vector field $u^a = e^{-\nu /2} \delta^a_0$ is not appropriate, since it gives $\rho = 0$ due to the condition $T_0 = 0$. Thus we use the spacelike four-velocity $u^a = e^{-\lambda/2} \delta^a_1$ of the observers for the metric \eqref{metric}, which implies that Eqs. \eqref{rho-p-emt-2} and \eqref{pi-123-emt-2} give
\begin{equation}
  \rho = e^{-\lambda} T_1 \, , \quad p = 0 \, , \quad \pi_{ab} = 0 \, .
\end{equation}
Thus, the dust fluid is also allowed in this subcase. Furthermore, all Ricci tensor components for this subcase are non-zero. This means that the Ricci tensor is non-degenerate even though the matter tensor is degenerate.

\bigskip

\noindent {\bf Case (D-A3)}. For this case, the possible subcases are given by {\bf (D-A3-i)} $T_{p} \neq 0, \,\, T_{q} = 0, \,\, (p = 0,1 \,\, {\rm and} \,\, q= 2,3)$ and {\bf (D-A3-ii)} $T_{p} = 0, \,\,
T_{q} \neq 0$.


{\bf Subcase (D-A3-i)}.  Here, by choosing a timelike four-velocity, the conditions $T_{0} \neq 0 \neq T_1$ and $T_{2} = 0 = T_3$ mean that the fluid represents an anisotropic fluid without heat flux, and then the physical quantities become
\begin{eqnarray}
& &  \rho = e^{-\nu} T_0 \, , \qquad p = \frac{1}{3} e^{- \lambda} T_1 \, , \label{rho-p-emt-DA3i} \\ & & \pi_{11} = 2 e^{\lambda} p  \, , \quad  \pi_{22} =  - p r^2 \, , \quad  \pi_{33} = - p  r^2 \sin^2 \theta \, . \label{pi-DA3i}
\end{eqnarray}
Using the transformations $d\bar{r} = \psi
\sqrt{\mid T_{1} \mid} dr$, where $\bar{r}=\bar{r}(t,r)$, one finds
\begin{equation}
X^0 = - \int{\frac{(\dot{\bar{r}}+\dot{f}) }{\psi T_{0}} g(t) d\bar{r}} , \qquad X^ 1 = \frac{\bar{r} + f(t)}{\sqrt{\mid T_{1} \mid}} , \qquad X^q = X^q (x^a),
\end{equation}
where $f(t)$ and $g(t)$ are functions of integration, and the conformal factor $\psi$ has the form
\begin{equation}
\psi = \frac{2 T_{0}}{(\bar{r} + f - 2T_{0})} \left[ \int{ \left(\frac{\dot{\bar{r}} + \dot{f}}{\psi} \right)^{.} \frac{d\bar{r}}{T_{0}}} - \dot{g}(t) \right].
\end{equation}
For this case, it follows from the condition $T_q =0$ that
\begin{eqnarray}
& & R_0 = \frac{e^{\nu- \lambda}}{2r}(\nu + \lambda)' \, , \qquad R_{1} = \frac{1}{2r}(\nu + \lambda)'  \, , \\ & & T_0 + e^{\nu -\lambda} T_1 = 2 R_0 \, , \qquad e^{-\nu} T_0 - e^{-\lambda} T_1 = \frac{2}{r^2} R_2 \, . \label{T0-T1-DA3i}
\end{eqnarray}
We find from Eqs. \eqref{rho-p-emt-DA3i} and \eqref{T0-T1-DA3i} that
\begin{equation}
\rho + 3 p = 2 e^{-\nu} R_0 \, , \qquad \rho - 3 p = \frac{2}{r^2} R_2 \, .
\end{equation}
These equations show that it should be satisfied the conditions $R_0 \geq e^{\nu} /2$ and $R_2 \geq 0$ in order to be valid the SEC and TEC, respectively.
There is only one constraint equation $ 2 \nu''+ {\nu'}^2- \nu' \lambda' + \frac{2}{r} (\nu' - \lambda')= 0$ following from the condition $T_2 = 0$ for this subcase. Also, all the Ricci tensor components are again non-zero, i.e. $\det ( R_{ab} ) \neq 0$,  even if the matter tensor is degenerate.


{\bf Subcase (D-A3-ii)}. For this subcase, considering the constraint $T_0 = 0 = T_1$, we find
\begin{equation}
\lambda' = \frac{1}{r} \left( 1- e^{\lambda} \right) = - \nu'  \, , \label{d-a3ii}
\end{equation}
which yields the following solution
\begin{equation}
  \nu (r) = \ln \left( 1 - \frac{\lambda_0}{r} \right) \, , \qquad  \lambda (r) = \ln \left( \frac{1}{1 - \frac{\lambda_0}{r}} \right) \, ,
\end{equation}
where $\lambda_0$ is an integration constant. The above solution is just the Schwarzschild metric which gives $R_{ab} = 0$, i.e., all $R_i$'s and $T_i$'s vanish identically.
Therefore there is a contradiction with the condition $T_2 \neq 0 \neq T_3$ as an assumption that is not possible in this subcase.

\bigskip

\noindent {\bf Case (D-A4)}. In this case, the possible subcases are {\bf
(D-A4-i)} $T_{0} = 0, \,\, T_{j} \neq 0, \,\, (j = 1,2,3)$ and
{\bf (D-A4-ii)} $T_{1} = 0, \,\, T_{k} \neq 0, \,\, (k = 0,2,3)$.


{\bf Subcase (D-A4-i)}. In this subcase, the constraint $T_0 =0$ gives
\begin{equation}
\lambda' = \frac{1}{r} \left( 1- e^{\lambda} \right) \, , \label{d-a4i1}
\end{equation}
and
\begin{eqnarray}
& & R_0 = - 2  e^{\nu} \left( e^{-\lambda} R_1 + \frac{1}{r^2} R_2 \right), \quad T_1 = -\frac{1}{2} \left( R_1 + \frac{4}{r^2}e^{\lambda} R_2 \right), \quad T_2 = - \frac{3}{2} r^2 e^{-\lambda} R_1 - R_2, \label{R012-DA4i}
\end{eqnarray}
where $T_1 = \frac{1}{r} \left( \nu + \lambda\right)'$. For this case the choice of a timelike four velocity of the observers is not allowed since $T_ 0 = 0$ gives $\rho = 0$. So we need to choose a spacelike four-velocity such as $u^a = e^{-\lambda/2} \delta^a_r$. Using Eqs. \eqref{rho-p-emt-2} and \eqref{pi-123-emt-2}, this choice gives rise to an anisotropic fluid without heat flux as follows:
\begin{eqnarray}
& &  \rho = e^{-\lambda} T_1 \, , \qquad p = \frac{2}{3 r^2} T_2 \, , \quad \pi_{00} = p e^{\lambda} \, , \quad \pi_{11} = 0 \, , \quad  \pi_{22} =  \frac{1}{2} p r^2 \, , \quad  \pi_{33} =  \sin^2 \theta \, \pi_{22}  \, . \label{rho-p-pi-emt-DA4i}
\end{eqnarray}
Here Eq. \eqref{d-a4i1} has the following solution
\begin{equation}
  \lambda (r) = \ln \left( \frac{1}{1 - \frac{\lambda_1}{r}} \right) \, , \label{lambda-DA4i}
\end{equation}
where $\lambda_1$ is a constant of integration. The equations given in \eqref{R012-DA4i} are second order ordinary differential equations in terms of $\nu$. Then, using the $\lambda(r)$ given by \eqref{lambda-DA4i} in any of the three equations of \eqref{R012-DA4i}, one can solve the obtained second order differential equation to find $\nu(r)$ as
\begin{equation}
   \nu (r) = 2 \ln \left\{ -\frac{\nu_0}{4} \sqrt{1 - \frac{\lambda_1 }{r} } + \frac{\nu_1}{2} \left[ r - 3 \lambda_1 + \frac{3}{2} \lambda_1 \sqrt{1 - \frac{\lambda_1 }{r} }  \ln \left( r + \sqrt{ r (r - \lambda_1)} - \frac{\lambda_1 }{2} \right) \right] \right\} \, , \label{nu-DA4i}
\end{equation}
where $\nu_0$ and $\nu_1$ are constants of integration. The Ricci scalar of the obtained metric given by \eqref{lambda-DA4i} and \eqref{nu-DA4i} is
\begin{equation}
  R = \frac{ 8 \nu_1}{ - \nu_0 \sqrt{r (r- \lambda_1)} + \nu_1 \left[ 2 r ( r-3 \lambda_1) + 3 \lambda_1 \sqrt{r (r- \lambda_1)} \ln \left( r + \sqrt{r (r- \lambda_1)} -\frac{\lambda_1}{2} \right) \right] } \, ,
\end{equation}
and the $R_i$'s and $T_i$'s for this solution are
\begin{equation}
  R_0 = \frac{1}{4 r^2} \left\{  - \nu_0 \sqrt{r (r- \lambda_1)} + \nu_1 \left[ 2 r ( r-3 \lambda_1) + 3 \lambda_1 \sqrt{r (r- \lambda_1)} \ln \left( r + \sqrt{r (r- \lambda_1)} -\frac{\lambda_1}{2} \right) \right] \right\} \, , \label{R0-DAi1}
\end{equation}
\begin{equation}
  R_1 = 0 \, , \, R_2 = \frac{ 2 \nu_1 r^2 }{ \nu_0 \sqrt{r (r- \lambda_1)} - \nu_1 \left[ 2 r ( r-3 \lambda_1) + 3 \lambda_1 \sqrt{r (r- \lambda_1)} \ln \left( r + \sqrt{r (r- \lambda_1)} -\frac{\lambda_1}{2} \right) \right] } \, , \label{R1R2-DAi1}
\end{equation}
\begin{equation}
T_1 = \frac{2}{ r ( \lambda_1 - r)} R_2  \, , \qquad T_2 = - R_2 \, . \label{T12-DAi1}
\end{equation}
Further, the matter density and pressure that come from Eqs. \eqref{rho-p-pi-emt-DA4i}, \eqref{lambda-DA4i} and \eqref{T12-DAi1} are
\begin{equation}
  \rho = \frac{2}{r^2} T_2 \, , \qquad p = \frac{1}{3} \rho \, ,
\end{equation}
which implies that the equation of state parameter $w$ yields an incohorent radiation, i.e., $w= 1/3$. Finally, we conclude that the Ricci tensor is also degenerate in this case.

For this subcase, we obtain that $X^0 = X^0 (x^a)$ and $X^j = X^j (r,\theta,\phi)$ where the form of $X^j$ is the same as in equations \eqref{x1} - \eqref{x3}, and the constraint equations \eqref{cnst-a} - \eqref{cnst-f} yield the following solution
\begin{equation}
A_j = b_j \cosh\bar{r} + d_j \sinh\bar{r},  \qquad A_5 = \ell \sqrt{ \frac{T_{2}}{T_{1}}},
\end{equation}
where $b_j, d_j$ and $\ell$ are constants. In this case, the conformal factor is given by
\begin{eqnarray}
& & \psi = \left( \frac{T_{{2},\bar{r}}}{2 T_{2}} A_{1,\bar{r}} + A_1 \right) \sin\theta \sin\phi - \left( \frac{T_{{2},\bar{r}}}{2 T_2} A_{2,\bar{r}} + A_2 \right) \sin\theta \cos\phi  \nonumber \\
& & \qquad + \left( \frac{T_{2,\bar{r}}}{2 T_2} A_{3,\bar{r}} + A_3 \right) \cos\theta + \ell \frac{T_{2,\bar{r}}}{2 T_2},
\end{eqnarray}
where we have used the transformation $dr = (T_{2}/T_{1})^{1/2} d\bar{r}$. By considering \eqref{R1R2-DAi1} and \eqref{T12-DAi1}, the latter transformation yields $\bar{r} = - \sqrt{2} \ln \left(  r + \sqrt{r (r- \lambda_1)} -\frac{\lambda_1}{2} \right)$. Thus, in addition to the three KVs given in \eqref{sphmin}, it follows that the remaining CMCs and corresponding conformal factors are
\begin{eqnarray}
& & \mathbf{X}_4= \sinh\bar{r} \cos \, \theta \partial_{\bar{r}} + \cosh\bar{r} \sin\theta \, \partial _\theta \, ,  \quad \psi_{4} = \cos\theta \left( \frac{T_{{2},\bar{r}}}{2 T_{2}} \sinh\bar{r} + \cosh\bar{r} \right) \, , \nonumber \\& & \mathbf{X}_5 = \cosh\bar{r} \cos \theta \, \partial_{\bar{r}} + \sinh\bar{r} \sin\theta \, \partial _\theta \, ,  \quad \psi_{5} = \cos\theta \left( \frac{T_{{2},\bar{r}}}{2 T_{2}} \cosh\bar{r} + \sinh\bar{r} \right) \, , \nonumber  \\& & \mathbf{X}_6 = \sinh\bar{r} \sin\theta \sin\phi \, \partial_{\bar{r}} - \cosh\bar{r} \, {\xi_1} \, , \quad \psi_{6} = \sin\theta \sin\phi \left( \frac{T_{{2},\bar{r}}}{2 T_{2}} \sinh\bar{r} + \cosh\bar{r} \right) \, , \nonumber \\& & \mathbf{X}_7 = \sinh\bar{r} \sin\theta \cos\phi \, \partial_{\bar{r}} - \cosh\bar{r} \, {\xi_2} \, , \quad \psi_{7} = \sin\theta \cos\phi \left( \frac{T_{{2},\bar{r}}}{2 T_{2}} \sinh\bar{r} + \cosh\bar{r} \right) \, , \label{a4i-cmc7} \\& & \mathbf{X}_8 = \cosh\bar{r} \sin\theta \sin\phi \, \partial_{\bar{r}} - \sinh\bar{r} \, {\xi_1} \, , \quad \psi_{8} = \sin\theta \sin\phi \left( \frac{T_{{2},\bar{r}}}{2 T_{2}} \cosh\bar{r} + \sinh\bar{r} \right) \, , \nonumber \\& & \mathbf{X}_9 = \cosh\bar{r} \sin\theta \cos\phi \, \partial_{\bar{r}} - \sinh\bar{r} \, {\xi_2} \, , \quad \psi_{9} = \sin\theta \cos\phi \left( \frac{T_{{2},\bar{r}}}{2 T_{2}} \cosh\bar{r} + \sinh\bar{r} \right) \, , \nonumber \\& & \mathbf{X}_{10} = \partial_{\bar{r}} \, , \quad \qquad \qquad \qquad \qquad \qquad \quad \,\,\,  \psi_{10} = \frac{T_{{2},\bar{r}}}{2 T_{2}} \, , \nonumber \\& & \mathbf{X}_{11} = F(t,r,\theta,\phi) \partial_{t} \, , \quad \qquad \qquad \qquad \quad \psi_{11} = 0 \, , \nonumber
\end{eqnarray}
where $F(t,r,\theta,\phi)$ is an arbitrary function, and we have
defined ${\xi}_1$ and ${\xi_2}$ as follows
\begin{eqnarray}
& & {\xi_1}=\cos\theta \sin\phi \, \partial _{\theta } + \csc\theta \cos\phi \, \partial _\phi \, ,  \qquad {\xi_2}= \cos\theta \cos\phi \, \partial_\theta - \csc\theta \sin\phi \, \partial_\phi \, .  \label{a4i-sphmin}
\end{eqnarray}
In order to construct a closed algebra for vector fields \eqref{a4i-cmc7}, we find that $F=F(t)$. Hence we have finite dimensional Lie algebra of CMCs which has the following non-vanishing commutators
\begin{eqnarray}
& & \left[\bf{X}_1,\bf{X}_2 \right] = \bf{X}_3, \qquad
\left[\bf{X}_1,\bf{X}_3 \right] = -\bf{X}_2, \quad
\left[\bf{X}_1,\bf{X}_4 \right] = -\bf{X}_6, \quad
\left[\bf{X}_1,\bf{X}_5 \right] = \bf{X}_8, \qquad
\left[\bf{X}_1,\bf{X}_6 \right] = \bf{X}_4, \nonumber \\& &
\left[\bf{X}_1,\bf{X}_8 \right] = \bf{X}_5, \qquad
\left[\bf{X}_2,\bf{X}_3 \right] = \bf{X}_1, \qquad
\left[\bf{X}_2,\bf{X}_4 \right] = -\bf{X}_7, \quad
\left[\bf{X}_2,\bf{X}_5 \right] = -\bf{X}_9, \quad
\left[\bf{X}_2,\bf{X}_7 \right] = \bf{X}_4, \nonumber \\& &
\left[\bf{X}_2,\bf{X}_9 \right] = \bf{X}_5, \qquad
\left[\bf{X}_3,\bf{X}_6 \right] = \bf{X}_7, \qquad
\left[\bf{X}_3,\bf{X}_7 \right] = -\bf{X}_6, \quad
\left[\bf{X}_3,\bf{X}_8 \right] = \bf{X}_9, \quad \,\,\,\,
\left[\bf{X}_3,\bf{X}_9 \right] = -\bf{X}_8, \nonumber \\& &
\left[\bf{X}_4,\bf{X}_5 \right] = -\bf{X}_{10}, \quad
\left[\bf{X}_4,\bf{X}_6 \right] = \bf{X}_1, \qquad
\left[\bf{X}_4,\bf{X}_7 \right] = \bf{X}_2, \quad \,\,\,
\left[\bf{X}_4,\bf{X}_{10} \right] = -\bf{X}_5, \,\,\,\,
\left[\bf{X}_5,\bf{X}_8 \right] = -\bf{X}_1, \label{Lie-alg-DA4i}  \\& &
\left[\bf{X}_5,\bf{X}_9 \right] = -\bf{X}_2, \,\, \quad
\left[\bf{X}_5,\bf{X}_{10} \right] = -\bf{X}_4, \quad
\left[\bf{X}_6,\bf{X}_7 \right] = -\bf{X}_3, \,\,\,
\left[\bf{X}_6,\bf{X}_8 \right] = -\bf{X}_{10},  \,\,\,\,
\left[\bf{X}_6,\bf{X}_{10} \right] = -\bf{X}_8, \nonumber \\& &
\left[\bf{X}_7,\bf{X}_9 \right] = -\bf{X}_{10}, \quad
\left[\bf{X}_7,\bf{X}_{10} \right] = -\bf{X}_9, \quad
\left[\bf{X}_8,\bf{X}_9 \right] = \bf{X}_3, \quad \,\,
\left[\bf{X}_8,\bf{X}_{10} \right] = -\bf{X}_6, \,\,\,\,
\left[\bf{X}_9,\bf{X}_{10} \right] = -\bf{X}_7. \nonumber
\end{eqnarray}


{\bf Subcase (D-A4-ii)}. For this subcase, where $T_1 = 0$, we have $\nu' = \frac{1}{r} \left( e^{\lambda} - 1 \right)$ and thus $T_0 = \frac{1}{r} e^{\nu -\lambda} \left( \lambda + \nu \right)'$. Also, $T_0$ and $T_2$ in terms of $R_i$'s ($i=0,1,2$) become
\begin{equation}
T_0 = R_0 + e^{\nu -\lambda} R_1  \, , \qquad T_2 = \frac{r^2}{2} \left( e^{-\nu} R_0 - e^{-\lambda} R_1 \right) \, ,
\end{equation}
and
\begin{equation}
  R_2 = \frac{r^2}{2} \left( e^{-\nu} R_0 + e^{-\lambda} R_1 \right) \, .
\end{equation}
In this subcase one can choose a timelike four-velocity of the observers such that $u^a = e^{-\nu} \delta^a_t$, which yields an anisotropic fluid without heat flux as
\begin{equation}
  \rho = e^{-\nu} T_0 \, , \quad p = \frac{2}{3 r^2} T_2 \, , \quad \pi_{11} = -e^{\lambda} p \, , \quad \pi_{22} \frac{r^2}{2} p \, , \quad \pi_{33} = \sin^2 \theta \, \pi_{22} \, .
\end{equation}
It is interesting to point out that we have a variable equation of state parameter $w = \frac{2}{3 \, r^2} e^{\nu}$, i.e., $p = \frac{2}{3 \, r^2} e^{\nu} \rho$ when $T_0 = T_2$.

\bigskip

If $T_{0} = T_{2}$, it follows from the constraint equations \eqref{cnst-a}-\eqref{cnst-f} that $A_j$ and $A_4$ have the following solutions
\begin{equation}
A_j = b_j \cosh\bar{r} + d_j \sinh\bar{r} \, , \qquad A_4 = \ell \, ,
\end{equation}
where $b_j, d_j$ and $\ell$ are integration constants. Then, the components of the CMC vector field are obtained as
\begin{eqnarray}
& & X^0 = \sin\theta \left[ A'_1 \sin\phi - A'_2 \cos\phi \right] + A'_3 \cos\theta + \ell \, , \nonumber \\& &  X^ 1 = \frac{2 T_{0}}{T'_{0}} \left[ \psi - \sin\theta(A_1 \sin\phi - A_2 \cos\phi) - A_3 \cos\theta \right] \, , \\& &  X^2 = -  \cos\theta (A_1 \sin\phi - A_2 \cos\phi) + A_3 \sin\theta + a_1 \sin\phi -
a_2 \cos\phi \, , \nonumber \\& & X^3 = - \csc \theta (A_1 \cos\phi + A_2 \sin\phi) + \cot\theta (a_1 \cos\phi + a_2 \sin\phi) + a_3 \, , \nonumber
\end{eqnarray}
where $T'_{0} \neq 0$, and $\psi$ is an arbitrary conformal factor, that is, the component $X^1$ is an arbitrary function of the coordinates and so we have \emph{infinite} dimensional algebra of CMCs. If $T_{0} \neq T_{2}$, then, in addition to the three KVs given in \eqref{sphmin}, we have the following CMCs
\begin{eqnarray}
& & \mathbf{X_4}= \sin\theta \sin\phi \, G(r) \, \partial_r -{\xi}_1 \, , \quad \,\,\,  \psi_{4} = \sin\theta \sin\phi \, G(r) \, , \nonumber \\& & \mathbf{X_5}= \sin\theta  \cos\phi \, G(r) \, \partial_{\bar{r}} + {\xi}_2 \, , \quad \,\,\, \psi_{5} = \sin\theta \cos\phi \, G(r) \, , \label{a4ii-cmc5} \\& & \mathbf{X_6}= \cos\theta \, G(r) \, \partial_{\bar{r}} + \sin\theta \, \partial_{\theta} \, , \quad \,\,\,\, \psi_{6} = \cos\theta \, G(r) \, ,  \nonumber \\& & \mathbf{X_7}= \dot{f}(t) \, G(r) \, \partial_{\bar{r}} + f(t) \, \partial_{t} \, , \qquad \,\, \psi_{7} = \dot{f} (t) \left[ 1- G(r) \right] \, , \nonumber
\end{eqnarray}
where $G(r)\equiv \frac{2 T_{0} /T_{2}}{(T_{0}/T_2)'}$, $(T_{0}/T_{2})' \neq 0$, and  $f(t)$ is an integration function. Thus, we again have an \emph{finite} dimensional Lie algebra of CMCs, and non-zero commutators of the Lie algebra have the following form:
\begin{eqnarray}
& & \left[\bf{X}_1,\bf{X}_2 \right] = \bf{X}_3 \, , \qquad
\left[\bf{X}_1,\bf{X}_3 \right] = -\bf{X}_2 \, , \,\, \quad
\left[\bf{X}_1,\bf{X}_4 \right] = \bf{X}_6, \quad \,\,\,\,\,
\left[\bf{X}_1,\bf{X}_6 \right] = -\bf{X}_4 \, , \nonumber \\& &
\left[\bf{X}_2,\bf{X}_3 \right] = \bf{X}_1 \, , \qquad
\left[\bf{X}_2,\bf{X}_5 \right] = \bf{X}_6 \, , \qquad
\left[\bf{X}_2,\bf{X}_6 \right] = -\bf{X}_5 \, , \quad
\left[\bf{X}_3,\bf{X}_4 \right] = \bf{X}_5 \, , \label{Lie-alg-DA4ii} \\& &
\left[\bf{X}_3,\bf{X}_5 \right] = -\bf{X}_4 \, , \quad
\left[\bf{X}_4,\bf{X}_5 \right] = -\bf{X}_3 \, , \quad \,\,
\left[\bf{X}_4,\bf{X}_6 \right] = -\bf{X}_1 \, , \quad
\left[\bf{X}_5,\bf{X}_6 \right] = -\bf{X}_2 \, . \nonumber
\end{eqnarray}

\bigskip

\section{Conformal Matter Collineations for the Non-degenerate Matter Tensor}
\label{non-degenerate}

\noindent In this section, we consider the CMCs in non-degenerate case, i.e. $\det (T_{ab}) \neq 0$, admitted by the static spherically symmetric spacetimes. Here we consider the following \emph{five}
possibilities of the non-degenerate matter tensor.

\bigskip

\noindent {\bf Case (ND-A)}. For this case, where none of the $T'_{a}\, (a= 0,1,2,3)$ is zero, applying the transformation $dr = \sqrt{T_{2} /T_{1}} d\bar{r}$, we find that the the number of CMCs is \emph{fifteen} such that there are three minimal KVs given in \eqref{sphmin}, and the remaining ones are
\begin{eqnarray}
& & {\bf X}_4 = f_1 (t) \left[ \sin\theta \sin\phi \,  h_1 (\bar{r}) {\bf Y} - h_2 (\bar{r}) \xi_1 \right] \, , \qquad \quad \,\,\, \psi_4 = f_1 (t) H_1 (\bar{r}) \sin\theta \sin\phi \, , \nonumber  \\ & &  {\bf X}_5 = f_2 (t) \left[ \sin\theta \sin\phi \,  h_1 (\bar{r}) {\bf Y} - h_2 (\bar{r}) \xi_1 \right] \, , \qquad \quad \,\,\, \psi_5 = f_2 (t) H_1 (\bar{r}) \sin\theta \sin\phi \, ,  \nonumber \\ & & {\bf X}_6 = \frac{1}{b} \left[ \sin\theta \sin\phi \,  h_2 (\bar{r}) {\bf Y} - h_1 (\bar{r}) \xi_1 \right] \, , \qquad \qquad \quad \psi_6 = \frac{H_2 (\bar{r})}{b} \sin\theta \sin\phi \, , \nonumber \\ & & {\bf X}_7 = f_1 (t) \left[ -\sin\theta \cos\phi \,  h_1 (\bar{r}) {\bf Y} + h_2 (\bar{r}) \xi_2 \right] \, , \qquad \,\,\, \psi_7 = - f_1 (t) H_1 (\bar{r}) \sin\theta \cos\phi \, , \nonumber  \\ & &  {\bf X}_8 = f_2 (t)  \left[ - \sin\theta \cos\phi \,  h_1 (\bar{r}) {\bf Y} + h_2 (\bar{r}) \xi_2 \right] \, , \qquad \,\,\, \psi_8 = - f_2 (t) H_1 (\bar{r}) \sin\theta \cos\phi \, ,  \nonumber \\ & & {\bf X}_9 = \frac{1}{b} \left[ -\sin\theta \cos\phi \,  h_2 (\bar{r}) {\bf Y} + h_1 (\bar{r})\xi_2 \right] \, , \qquad \qquad  \psi_9 = -\frac{H_2 (\bar{r})}{b} \sin\theta \cos\phi \, , \label{cmc-NDA} \\ & & {\bf X}_{10} = f_1 (t) \left[ \cos\theta  \,  h_1 (\bar{r}) {\bf Y} + h_2 (\bar{r}) \sin\theta \, \partial_{\theta} \right] \, , \qquad \quad \psi_{10} = f_1 (t) H_1 (\bar{r}) \cos\theta  \, , \nonumber  \\ & &  {\bf X}_{11} = f_2 (t) \left[ \cos\theta  \,  h_1 (\bar{r}) {\bf Y} + h_2 (\bar{r}) \sin\theta \, \partial_{\theta} \right] \, , \qquad \quad \psi_{11} = f_2 (t) H_1 (\bar{r}) \cos\theta \, ,  \nonumber \\ & & {\bf X}_{12} = \frac{1}{b} \left[ \cos\theta \,  h_2 (\bar{r}) {\bf Y} + h_1 (\bar{r}) \sin\theta \, \partial_{\theta} \right] \, , \qquad \qquad \,\,\, \psi_{12} = \frac{H_2 (\bar{r})}{b} \cos\theta \, , \nonumber \\ & & {\bf X}_{13} = - \ddot{f}_1 \frac{ \sinh \bar{r} }{a \, h_2 (\bar{r}) } \partial_t + \dot{f}_1 \, \partial_{\bar{r}} \, , \qquad \qquad  \qquad \qquad \quad \, \psi_{13} = \dot{f}_1 \frac{ T_{2,\bar{r}}}{ 2 \sqrt{T_1 T_2}} \, , \nonumber \\ & & {\bf X}_{14} = - \ddot{f}_2 \frac{ \sinh \bar{r} }{a \, h_2 (\bar{r}) } \partial_t + \dot{f}_2 \, \partial_{\bar{r}} \, , \qquad \qquad  \qquad \qquad \quad \,  \psi_{14} =  \dot{f}_2 \frac{ T_{2,\bar{r}}}{ 2 \sqrt{T_1 T_2}} \, , \nonumber \\ & & {\bf X}_{15} =  \partial_t \, , \qquad \qquad  \qquad \qquad \qquad \qquad \qquad \qquad \quad \, \psi_{15} = 0 \, , \nonumber
\end{eqnarray}
where $H_1 (\bar{r}), H_2 (\bar{r}), h_1 (\bar{r}), h_2 (\bar{r}), f_1 (t), f_2 (t)$ and ${\bf Y}$ are defined as follows:
\begin{eqnarray}
& & H_1 (\bar{r}) = h_1 (\bar{r})  \frac {T_1 T_{2,\bar{r}}}{2 T_2^{2}} + h_2 (\bar{r})  \, , \quad H_2 (\bar{r}) = h_2 (\bar{r}) \frac {T_1 T_{2,\bar{r}}}{2 T_2^{2}} + h_1 (\bar{r}) \, , \\ & & h_1 (\bar{r}) = a \sinh \bar{r} + b \cosh \bar{r} \, , \qquad h_2 (\bar{r}) = a \cosh \bar{r} + b \sinh \bar{r} \, , \\ & & f_1 (t) = \left\{
\begin{array}{l}
\frac{1}{\alpha} \sinh( \alpha t) \, , \quad \,\,\,
{\rm for} \,\, \alpha^2 > 0, \\
\frac{1}{ |\alpha| } \sin( |\alpha| t) \, , \quad {\rm for} \,\,
\alpha^2 < 0,
\end{array}
\right.  \label{f1-cmc} \\ & & f_2 (t) = \left\{
\begin{array}{l}
\frac{1}{\alpha} \cosh( \alpha t) \, , \qquad \,\,  {\rm for} \,\,
\alpha^2 > 0, \\ -\frac{1}{|\alpha|} \cos( |\alpha| t) \, , \quad
{\rm for} \,\, \alpha^2 < 0,
\end{array}
\right.  \label{f2-cmc} \\ & & {\bf Y} = \frac{ \sqrt{T_1 T_2}}{T_0} \partial_t + \sqrt{\frac{T_2}{T_1}} \partial_{r} = \frac{ \sqrt{T_1 T_2}}{T_0} \partial_t + \partial_{\bar{r}} \, , \label{cmc-Y}
\end{eqnarray}
and $T_{0} = h_2 (\bar{r})^2 T_{2}$\, , $a$ and $b$ are integration constants, and $\alpha$ is a constant of separation such that
\begin{equation}
  \alpha^2 = \left\{
\begin{array}{l}
a^2 - b^2 \, , \quad {\rm for} \,\, \alpha^2 > 0\, , \\
b^2 - a^2 \, , \quad {\rm for} \,\, \alpha^2 < 0\, .
\end{array}
\right.
\end{equation}

\noindent For $\alpha^2 = 0$, after solving the constraint equations \eqref{cnst-a}-\eqref{cnst-f}, we find \emph{twelve} CMCs as follows: the KVs ${\bf X}_1, {\bf X}_2, {\bf X}_3$ given in \eqref{sphmin}, and
\begin{eqnarray}
& & {\bf X}_4 = e^{\beta \bar{r}} \left[ \beta\, \sin\theta \sin\phi \, {\bf Y} - \xi_1 \right] \, , \qquad \psi_4 =  H (\bar{r}) e^{\beta \bar{r}} \sin\theta \sin\phi \, ;  \qquad  {\bf X}_5 = c_0 \, t \, {\bf X}_4 \, , \quad  \psi_5 = c_0 \, t \, \psi_4 \, ;  \nonumber \\ & & {\bf X}_6 = e^{\beta \bar{r}} \left[ - \beta \,  \sin\theta \cos\phi \,  {\bf Y} +  \xi_2 \right] \, , \quad \psi_6 = - e^{\beta \bar{r}} H (\bar{r}) \sin\theta \cos\phi \, ; \quad {\bf X}_7 = c_0 \, t \,{\bf X}_6 \, , \quad \psi_7 = c_0 \, t \, \psi_6 \, ; \nonumber \\ & & {\bf X}_8 = e^{\beta \bar{r}} \left[ \beta \, \cos\theta  \,  {\bf Y} + \sin\theta \, \partial_{\theta} \right] \, , \qquad \psi_8 = e^{\beta \bar{r}} H (\bar{r}) \cos\theta  \, ; \qquad   {\bf X}_9 = c_0 \, t \, {\bf X}_8  \, , \quad \psi_9 = c_0 \, t \, \psi_8 \, ;  \label{cmc-NDA-ii} \\ & & {\bf X}_{10} = \frac{1}{2} \left( \frac{e^{-2 \beta \bar{r}}}{\beta \, c_0^2}   - \beta \, t^2 \right) \partial_t + t \,  \partial_{\bar{r}} \, , \qquad \psi_{10} = \frac{ t \, T_{2,\bar{r}}}{ 2 \sqrt{T_1 T_2}} \, , \nonumber \\ & & {\bf X}_{11} = - \beta t \, \partial_t + \partial_{\bar{r}} \, , \quad   \psi_{11} =  \frac{ T_{2,\bar{r}}}{ 2 \sqrt{T_1 T_2}} \, ; \quad {\bf X}_{12} =  \partial_t \, , \quad \ \psi_{12} = 0 \, , \nonumber
\end{eqnarray}
where $H (\bar{r})$ is defined as
\begin{equation}
  H (\bar{r}) = \beta  \frac {T_1 T_{2,\bar{r}}}{2 T_2^{2}}  + 1 \, ,
\end{equation}
and $T_{0} = c^2_0 e^{2 \beta \bar{r}} T_{2}$, $T_1 = (c_1 e^{-2 \beta \bar{r}} + \beta) T_2 $, $\beta$ is a separation constant such that $\beta = \pm 1$, $c_0$ and $c_1$ are integration constants.

\bigskip

\noindent {\bf Case (ND-B)}. Three of the $T'_{a}$ are zero. In this case we have the possibilities: {\bf (ND-B-i)} $T'_{0} \neq 0, \,\, T'_{j} = 0, \,\, (j = 1,2,3)$ and {\bf (ND-B-ii)} $T'_{1} \neq 0, \,\,
T'_{k} = 0, \,\, (k = 0,2,3)$.

{\bf Subcase (ND-B-i).} For this subcase we have 15 CMCs which are the same form as \eqref{cmc-NDA} together with the KVs ${\bf X}_1, {\bf X}_2, {\bf X}_3$ given in \eqref{sphmin}, under the transformations $\bar{r} \rightarrow k \, r $, $h_1 (\bar{r}) \rightarrow h_1 (r) = a \sinh (k r) + b \cosh (k r)$, $h_2 (\bar{r}) \rightarrow h_2 (r) = a \cosh (k r) + b \sinh (k r)$, $H_1 (\bar{r}) \rightarrow h_2 (r)$ and $H_2 (\bar{r}) \rightarrow h_1 (r)$, where $k = c_1 /c_2$, and $a, b, c_1, c_2$ are non-zero constants, and
\begin{eqnarray}
& & T_1 = \pm c_1^2, \quad \,\,  T_2= \pm c_2^2, \quad \,\, \alpha^2 = \pm (a^2 - b^2)/c_2^2 \, ,  \qquad  T_0 = \left[ a \cosh (k\, r)  + b \sinh (k\, r ) \right]^2 \, .
\end{eqnarray}
Here we note that the vector fields ${\bf X}_{13}, {\bf X}_{14}$ and ${\bf X}_{15}$ are MCs since the scale factors for those are zero, i.e., $\psi_{13}=0, \psi_{14}=0$ and $\psi_{15}=0$.

When $\alpha^2 = 0$, one finds that there are 15 CMCs which are the KVs ${\bf X}_1, {\bf X}_2, {\bf X}_3$ given in \eqref{sphmin} and
\begin{eqnarray}
& & {\bf X}_4 = \beta \, K_{_{-}} (t,r) \, \sin\theta \sin\phi \, {\bf Y} - K_{+} (t,r) \, \xi_1  \, , \qquad \quad \,\,\, \psi_4 = K_{+} (t,r) \sin\theta \sin\phi \, , \nonumber  \\ & &  {\bf X}_5 = \frac{c_0 \, \beta}{c_1} e^{\beta \, r} t \left[ \sin\theta \sin\phi \,  {\bf Y} - \xi_1 \right] \, , \qquad \qquad \qquad \quad  \psi_5 = \frac{c_0 \, \beta}{c_1} e^{\beta \, r} t \, \sin\theta \sin\phi \, ,  \nonumber \\ & & {\bf X}_6 = e^{\beta \, r} \left[ \sin\theta \sin\phi \, {\bf Y} - \xi_1 \right] \, , \qquad \qquad \qquad \qquad \quad \, \psi_6 = e^{\beta \, r} \sin\theta \sin\phi \, , \nonumber \\ & & {\bf X}_7 =  - \beta \, K_{_{-}} (t,r) \sin\theta \cos\phi \, {\bf Y} + K_{+} (t,r)\xi_2  \, , \qquad \,\,\,\,\,  \psi_7 = - K_{+} (t,r) \sin\theta \cos\phi \, , \nonumber  \\ & &  {\bf X}_8 = \frac{c_0 \, \beta}{c_1} e^{\beta \, r} t  \left[ - \sin\theta \cos\phi \,  {\bf Y} +  \xi_2 \right] \, , \qquad \qquad \quad \,\,\,\,\,  \psi_8 = - \frac{c_0 \, \beta}{c_1} e^{\beta \, r} t \sin\theta \cos\phi \, ,  \nonumber \\ & & {\bf X}_9 = e^{\beta \, r} \left[ -\sin\theta \cos\phi \,  {\bf Y} +  \xi_2 \right] \, , \qquad \qquad \qquad \quad \,\,\,\,\,\,  \psi_9 = - e^{\beta \, r} \sin\theta \cos\phi \, , \label{cmc-NDBib} \\ & & {\bf X}_{10} = \beta \, K_{_{-}} (t,r) \cos\theta  \,   {\bf Y} + K_{+} (t,r) \sin\theta \, \partial_{\theta}  \, , \qquad \quad   \psi_{10} = K_{+} (t,r) \cos\theta  \, , \nonumber  \\ & &  {\bf X}_{11} = \frac{c_0 \, \beta}{c_1} e^{\beta \, r} t  \left[ \cos\theta  \,  {\bf Y} + \sin\theta \, \partial_{\theta} \right] \, , \qquad \qquad \qquad \,\, \psi_{11} = \frac{c_0 \, \beta}{c_1} e^{\beta \, r} t \, \cos\theta \, ,  \nonumber \\ & & {\bf X}_{12} = e^{\beta \, r} \left[ \cos\theta \, {\bf Y} + \sin\theta \, \partial_{\theta} \right] \, , \qquad \qquad \qquad \qquad \,\,\, \psi_{12} = e^{\beta \, r} \cos\theta \, , \nonumber \\ & & {\bf X}_{13} =  \frac{1}{2} \left( \frac{T_1 \, e^{-2\beta \, r}}{\beta\, c_0^2} - \beta \, t^2 \right) \partial_t + t \, \partial_{r} \, , \qquad \qquad \quad \,\, \psi_{13} = 0 \, , \nonumber \\ & & {\bf X}_{14} = - \beta \, t \,  \partial_t +  \partial_{r} \, , \qquad \qquad  \qquad \qquad \qquad \qquad \,\,\,\,\,  \psi_{14} =  0 \, , \nonumber \\ & & {\bf X}_{15} =  \partial_t \, , \qquad \qquad  \qquad \qquad \qquad \qquad \qquad \qquad \quad \, \psi_{15} = 0 \, , \nonumber
\end{eqnarray}
where $\beta$ is an integration constant, $T_0 = c_0^2 e^{2 \beta r},  T_1 = \pm c_1^2, \beta^2 = T_1 / T_2 $, ${\bf Y}$ is given in \eqref{cmc-Y}, and $K_{\pm} (t,r)$ is defined as
\begin{equation}
  K_{\pm} (t,r) = \frac{1}{2} \left( \frac{c_0 \, \beta}{c_1} e^{\beta \, r} t^2 \pm \frac{c_1}{c_0 \, \beta} e^{-\beta \, r} \right) \, .
\end{equation}
It is explicitly seen from \eqref{cmc-NDBib} that in addition to the KVs ${\bf X}_1, {\bf X}_2, {\bf X}_3$ given in \eqref{sphmin}, the vector fields ${\bf X}_{13}, {\bf X}_{14}$ and ${\bf X}_{15}$ are MCs.


{\bf Subcase (ND-B-ii)}. In this case, where $T_0, T_2$ and $T_3$ are constants, one can easily find that the number of CMCs is \emph{six}, and these reduce to the MCs which are given by three KVs ${\bf X}_1, {\bf X}_2, {\bf X}_3$ given in \eqref{sphmin}, and the remaining ones
\begin{eqnarray}
& & {\bf X}_{4} = \partial_t \, , \qquad {\bf X}_{5} = \partial_{\bar{r}} \, , \qquad {\bf X}_{6} = \bar{r} \partial_t - t \partial_{\bar{r}} \, ,
\end{eqnarray}
where we have used the rescaling $d\bar{r} = \sqrt{T_{1}} dr$.

\bigskip

\noindent {\bf Case (ND-C)}. Two of the $T'_a$ are zero. In this case, the possible subcases are {\bf (ND-C-i)} $\ T'_{p} \neq 0, \,\, T'_{q} = 0, \,\, (p = 0,1 \,\, {\rm and} \,\, q= 2,3)$ and {\bf
(ND-C-ii)} $ T'_{p} = 0, \,\, T'_{q} \neq 0$.


{\bf Subcase (ND-C-i)}. For this subcase, where $T_2, T_3$ are constants, we find 15 CMCs that are similar to the ones given in \eqref{cmc-NDA}. Here, the functions $f_1(t)$ and $f_2(t)$ are respectively the same form given in Eqs. \eqref{f1-cmc} and \eqref{f2-cmc}, and the vector field ${\bf Y}$ has the form
${\bf Y} = \frac{\sqrt{T_1 }}{T_0} \partial_t + \partial_{\bar{r}}$. Also, the functions $ h_1 (\bar{r}), h_2 (\bar{r}), H_1 (\bar{r})$ and $H_2 (\bar{r})$ in this subcase have the following form
\begin{eqnarray}
& & h_1 (\bar{r}) = a \sinh \left(  \frac{\bar{r}}{c_2} \right) + b \cosh \left(  \frac{\bar{r}}{c_2} \right) \, , \quad h_2 (\bar{r}) = \frac{1}{c_2} \left[ a \cosh \left(  \frac{\bar{r}}{c_2} \right) + b \sinh \left(  \frac{\bar{r}}{c_2} \right) \right] \, , \\ & & H_1 (\bar{r}) =  h_2 (\bar{r})  \, , \quad H_2 (\bar{r}) = h_1 (\bar{r}) \, ,
\end{eqnarray}
with $dr = d\bar{r} / \sqrt{{|T_{1}|}} , T_0 = T_2 \, h_ 2 (\bar{r})^2 \, , T_2 = c_2^2$ and $\alpha^2 = \pm (a^2 -b^2)/T_2$.  The CMCs ${\bf X}_{13}, {\bf X}_{14}$ and ${\bf X}_{15}$ of this subcase reduce to MCs since the scale factors of these vector fields are zero.

When $\alpha^2 = 0$, there are twelve CMCs which are in the form \eqref{cmc-NDA-ii}, by replacing $\beta {\bf Y} \rightarrow {\bf Y} = \frac{\sqrt{T_1 }}{T_0} \partial_t + \partial_{\bar{r}}$ and $H (\bar{r}) \rightarrow 1$. For this subcase, the CMCs  ${\bf X}_{10}, {\bf X}_{11}$ and  ${\bf X}_{12}$ become MCs due to $\psi_{10} = 0 = \psi_{11}$ and $\psi_{12} = 0$.


{\bf Subcase (ND-C-ii)}. In this subcase, where $T_0, T_1$ are constants, there appears a constant of separation $\alpha$ that is given by the following constraint equation
\begin{equation}
   \frac{T_0}{2 T_1 \sqrt{T_2}} \left( \frac{T_2'}{\sqrt{T_2}} \right)'  = \alpha^2 \, . \label{const-alpha-ND-Cii}
\end{equation}
Then, it follows for $\alpha^2 \neq 0$ that there are 15 CMCs such that the KVs ${\bf X}_1, {\bf X}_2, {\bf X}_3$ given in \eqref{sphmin} and
\begin{eqnarray}
& & {\bf X}_4 = - \sqrt{T_0}\, f_3 (t) \left( \frac{T_{2,\bar{r}}}{2}  \sin\theta \sin\phi \,  {\bf Z} + \frac{1}{\sqrt{T_2}} \xi_1 \right) \, , \quad \,\,\, \psi_4 = f_3 (t) H_3 (\bar{r}) \sin\theta \sin\phi \, , \nonumber  \\ & &  {\bf X}_5 = -\sqrt{T_0} \, f_4 (t) \left( \frac{T_{2,\bar{r}}}{2} \sin\theta \sin\phi \,  {\bf Z} +  \frac{1}{\sqrt{T_2}} \xi_1 \right) \, , \quad \,\,\, \psi_5 = f_4 (t) H_3 (\bar{r}) \sin\theta \sin\phi \, ,  \nonumber \\ & & {\bf X}_6 = - T_1 \sqrt{b} \sin\theta \sin\phi \, {\bf Z} - \frac{b T_{2,\bar{r}} }{\sqrt{T_2}} \xi_1  \, , \qquad \qquad \qquad \quad  \psi_6 = H_4 (\bar{r}) \sin\theta \sin\phi \, , \nonumber \\ & & {\bf X}_7 = \sqrt{T_0}\, f_3 (t) \left( \frac{T_{2,\bar{r}}}{2} \sin\theta \cos\phi \,  {\bf Z} + \frac{1}{\sqrt{T_2}} \xi_2 \right) \, , \qquad \,\,\, \psi_7 = - f_3 (t) H_3 (\bar{r}) \sin\theta \cos\phi \, , \nonumber  \\ & &  {\bf X}_8 = \sqrt{T_0}\, f_4 (t)  \left( \frac{T_{2,\bar{r}}}{2} \sin\theta \cos\phi \, {\bf Z} + \frac{1}{\sqrt{T_2}} \xi_2 \right) \, , \qquad \,\,  \psi_8 = - f_4 (t) H_3 (\bar{r}) \sin\theta \cos\phi \, ,  \nonumber \\ & & {\bf X}_9 = T_1 \sqrt{b} \sin\theta \cos\phi \,  {\bf Z} + \frac{b T_{2,\bar{r}}}{\sqrt{T_2}} \xi_2 \, , \qquad \qquad \qquad \quad \,\,\,\, \psi_9 = - H_4 (\bar{r}) \sin\theta \cos\phi \, , \label{cmc-NDCii-a} \\ & & {\bf X}_{10} = \sqrt{T_0}\, f_3 (t) \left( - \frac{T_{2,\bar{r}}}{2} \cos\theta  \,  {\bf Z} + \frac{\sin\theta}{\sqrt{T_2}} \, \partial_{\theta} \right) \, , \qquad \,\,\,\, \psi_{10} = f_3 (t) H_3 (\bar{r}) \cos\theta  \, , \nonumber  \\ & &  {\bf X}_{11} = \sqrt{T_0} \, f_4 (t) \left( -\frac{T_{2,\bar{r}}}{2} \cos\theta  \,  {\bf Z} + \frac{\sin\theta}{\sqrt{T_2}} \, \partial_{\theta} \right) \, , \qquad \,\,\,\, \psi_{11} = f_4 (t) H_3 (\bar{r}) \cos\theta \, ,  \nonumber \\ & & {\bf X}_{12} = - T_1 \sqrt{b} \cos\theta \, {\bf Z} + \frac{ b T_{2, \bar{r}}}{\sqrt{T_2}} \sin\theta \, \partial_{\theta} \, , \qquad \qquad \qquad  \psi_{12} = H_4 (\bar{r}) \cos\theta \, , \nonumber \\ & & {\bf X}_{13} = \left( \frac{a}{2} f_1 - \frac{T_1}{T_0} \ddot{f}_1 \bar{r} \right) \partial_t + T_2 \dot{f}_1 \, \partial_{\bar{r}} \, , \qquad \qquad  \qquad \,\,\,\,\, \psi_{13} = \dot{f}_1 \frac{ T_{2,\bar{r}}}{2} \, , \nonumber \\ & & {\bf X}_{14} = \left( \frac{a}{2} f_2 - \frac{T_1}{T_0} \ddot{f}_2 \bar{r} \right) \partial_t + T_2 \dot{f}_2 \, \partial_{\bar{r}} \, , \qquad \qquad  \qquad \,\,\,\,\,  \psi_{14} =  \dot{f}_2 \frac{ T_{2,\bar{r}}}{ 2} \, , \nonumber \\ & & {\bf X}_{15} =  \partial_t \, , \qquad \qquad  \qquad \qquad \qquad \qquad \qquad \qquad \qquad \, \psi_{15} = 0 \, , \nonumber
\end{eqnarray}
where $H_3 (\bar{r}), H_4 (\bar{r}), f_3 (t), f_4 (t)$ and ${\bf Z}$ are defined as follows:
\begin{eqnarray}
& & H_3 (\bar{r}) = \sqrt{ \frac{T_0}{T_2}} \left( 1-  \frac { T_{2,\bar{r}}^2 }{4 T_2} \right)  \, , \quad H_4 (\bar{r}) = \left( b - \frac{ T_1 \sqrt{b}}{T_2}  \right) \frac { T_{2,\bar{r}}}{ \sqrt{T_2}}  \, , \\ & & f_3 (t) = \left\{
\begin{array}{l}
\frac{1}{\alpha} \sin ( \alpha t) \, , \qquad \,\,\,\,
{\rm for} \,\, \alpha^2 > 0, \\
\frac{1}{ |\alpha| } \sinh ( |\alpha| t) \, , \quad {\rm for} \,\,
\alpha^2 < 0,
\end{array}
\right.  \label{f1-cmc-ii} \\ & & f_4 (t) = \left\{
\begin{array}{l}
\frac{1}{\alpha} \cos ( \alpha t) \, , \qquad \quad \,\,\,  {\rm for} \,\,
\alpha^2 > 0, \\ - \frac{1}{|\alpha|} \cosh ( |\alpha| t) \, , \quad
{\rm for} \,\, \alpha^2 < 0,
\end{array}
\right.  \label{f2-cmc-ii} \\ & & {\bf Z} =  \frac{1}{T_0} \partial_t + \frac{\sqrt{T_2}}{T_1} \partial_{\bar{r}} \, . \label{cmc-Z}
\end{eqnarray}
Here, Eq. \eqref{const-alpha-ND-Cii} becomes
\begin{equation}
   \frac{T_0 T_{2,\bar{r} \bar{r}} }{2 T_1} = \alpha^2 \, , \label{const-alpha-2-ND-Cii}
\end{equation}
by using the transformation $dr= d\bar{r} / \sqrt{|T_{2}|}$, and it has a solution  $T_{2} = \frac{\alpha^2 T_1}{T_0} \bar{r}^2 + a\, \bar{r} + b$, where $a$ and $b$ are integration constants. Further, we have a relation $T_0 = 4 b \, \alpha^2 T_1 /(a^2 - 4 T_1)$ from the constraint equation \eqref{cnst-f}.

If we consider the possibility $\alpha^2 = 0$ in \eqref{const-alpha-2-ND-Cii}, it yields $T_2 = a\, \bar{r} + b$. Then one can find 15 CMCs which are three KVs ${\bf X}_1, {\bf X}_2, {\bf X}_3$ given in \eqref{sphmin}, and the following proper CMCs
\begin{eqnarray}
& & {\bf X}_4 = - \sqrt{T_0}\, t \left( \frac{a}{2}  \sin\theta \sin\phi \,  {\bf Z} + \frac{1}{\sqrt{T_2}} \xi_1 \right) \, , \,\,\, \psi_4 = t H_3 (\bar{r}) \sin\theta \sin\phi \, ;\,\, {\bf X}_5 = \frac{t}{2} {\bf X}_4 \, , \,\, \psi_5 = \frac{t}{2} \psi_4 \, \nonumber  \\ & & {\bf X}_6 = - \frac{1}{2 a} \sin\theta \sin\phi \, {\bf Z} - \frac{1}{\sqrt{T_2}} \xi_1  \, , \qquad \qquad  \psi_6 = H_5 (\bar{r}) \sin\theta \sin\phi \, , \nonumber \\ & & {\bf X}_7 = \sqrt{T_0}\, t \left( \frac{a}{2} \sin\theta \cos\phi \,  {\bf Z} + \frac{1}{\sqrt{T_2}} \xi_2 \right) \, , \,\,\, \psi_7 = - t H_3 (\bar{r}) \sin\theta \cos\phi \, ; \,\, {\bf X}_8 = \frac{t}{2} {\bf X}_7 \, , \,\, \psi_8 = \frac{t}{2} \psi_7 \, ,  \nonumber  \\ & & {\bf X}_9 = \frac{1}{2 a} \sin\theta \cos\phi \,  {\bf Z} + \frac{1}{\sqrt{T_2}} \xi_2 \, , \qquad \qquad   \psi_9 = - H_5 (\bar{r}) \sin\theta \cos\phi \, , \label{cmc-NDCii-b} \\ & & {\bf X}_{10} = \sqrt{T_0}\, t \left( - \frac{a}{2} \cos\theta  \,  {\bf Z} + \frac{\sin\theta}{\sqrt{T_2}} \, \partial_{\theta} \right) \, , \,\,\,  \psi_{10} = t H_3 (\bar{r}) \cos\theta  \, ;\,\, {\bf X}_{11} =  \frac{t}{2} {\bf X}_{10} \, , \,\, \psi_{11} = \frac{t}{2} \psi _{10} \, , \nonumber \\ & & {\bf X}_{12} = - \frac{1}{2 a} \cos\theta \, {\bf Z} + \frac{\sin\theta}{\sqrt{T_2}}  \, \partial_{\theta} \, , \qquad \qquad  \psi_{12} = H_5 (\bar{r}) \cos\theta \, , \nonumber \\ & & {\bf X}_{13} = \frac{a}{2}  \left( \frac{t^2}{2} - \frac{\bar{r}}{\sqrt{T_0}} \right) \partial_t + \frac{2 T_2}{a} t \, \partial_{\bar{r}} \, , \quad \psi_{13} = t \, , \nonumber \\ & & {\bf X}_{14} = t\, \partial_t + \frac{2 T_2}{a} \, \partial_{\bar{r}} \, , \quad \psi_{14} =  1 \, ; \,\, {\bf X}_{15} =  \partial_t \, , \,\,\, \psi_{15} = 0 \, , \nonumber
\end{eqnarray}
where $r = \frac{2}{a} \sqrt{ a \, \bar{r} + b}, T_1 = a^2 /4$, and $H_5 ({\bar{r}})$ is defined by
\begin{eqnarray}
& & H_5 (\bar{r}) = \frac{1}{\sqrt{T_2}} \left( 1 - \frac{a^3}{4 T_2} \right) \ .
\end{eqnarray}

\bigskip

\noindent {\bf Case (ND-D)}. One of the $T'_{a}$ is zero. In this case, the possibilities are {\bf (ND-D-i)} $ T'_{0} = 0, \,\, T'_{j} \neq 0$,  $(j = 1,2,3)$ and {\bf (ND-D-ii)} $\ T'_{1} = 0, \,\, T'_{k} \neq 0, \,\, (k = 0,2,3)$.

In the subcase {\bf (ND-D-i)}, if $\alpha^2 \neq 0$, then one obtains 15 CMCs as given by \eqref{sphmin} and \eqref{cmc-NDA} in the case {\bf (ND-A)}, in which the differences are the conditions $T_0 = c_0^2$, $T_2 = [ a \cosh \bar{r} + b \sinh \bar{r}]^{-2}$, and $\alpha^2 = \pm T_0 (a^2 - b^2)$ which is a separation constant. Also, when $\alpha^2 = 0$, we find 12 CMCs which are the same form as \eqref{cmc-NDA-ii} in the case {\bf (ND-A)}, where the differences are $T_0 = c_0^2$ and $T_2 = c_2^2 e^{-2 \beta \bar{r}}$.

For the subcase {\bf (ND-D-ii)}, we again find 15 CMCs for $\alpha^2 \ne 0$, which are the same form as given in \eqref{sphmin} and \eqref{cmc-NDA}, and the differences come from the constraints as $T_1 = c_1^2, T_0 = T_2 \left[ a \cosh \bar{r} + b \sinh \bar{r}  \right]^2$, and $\alpha^2 = \pm ( a^2 -b^2 )/T_1$. Further, if $\alpha^2 = 0$, then we find 12 CMCs which are the same form as in \eqref{sphmin} and \eqref{cmc-NDA-ii} together with the constraints $T_1 = c_1^2, T_0 = c_0^2 T_2 e^{2 \beta \bar{r}}$ and $\beta^2 =1$.

\bigskip

\noindent {\bf Case (ND-E)}. All $T'_{a}$ are zero. In this case, the constraints are $T_{0} = c_0, \, T_{1} = c_1, \, T_{2} = c_{2}$, where $c_0, c_1$ and $c_{2}$ are non-zero constants. Using these constraints, it follows that in addition to the three KVs given in \eqref{sphmin} there are three additional CMCs such as
\begin{equation}
{\bf X}_4 = t \, \partial_r - \frac{c_1}{c_0} r \, \partial_t  \, , \qquad {\bf X}_5 = \partial_r \, , \qquad {\bf X}_6 = \partial_t \, ,
\end{equation}
with $\psi_4, \psi_5, \psi_6 = 0$,  which means that these CMCs reduce to MCs.

Here, using the constraints $T_{0} = c_0, \, T_{1} = c_1, \, T_{2} = c_{2}$ in Eqs. \eqref{t0}, \eqref{t1} and \eqref{t2}, we have the following ordinary differential equations:
\begin{eqnarray}
& & \lambda' = \frac{1}{r} \left( 1 - e^{\lambda} \right) + c_0 \, r \, e^{\lambda - \nu} \, , \label{nde-t0} \\ & &  \nu' = \frac{1}{r} \left( e^{\lambda} -1 \right) + c_1 \, r \, , \label{nde-t1} \\ & & 2 \nu'' + \nu'^2 - \nu' \lambda' + \frac{2}{r} ( \nu' - \lambda') - \frac{4}{r^2} c_2 e^{\lambda} = 0 \, . \label{nde-t2}
\end{eqnarray}
Then, putting $\lambda'$ and $\nu'$ given above into \eqref{nde-t2}, yields
\begin{equation}
  e^{ 2 \lambda - \nu} - (c_1 \, r^2 + 1) e^{\lambda - \nu} + ( 3 c_1 \, r^2 - 4 c_2 ) \frac{ e^{\lambda}}{c_0\, r^2} + \frac{c_1}{c_0} ( c_1 r^2 + 1) = 0 \, ,
\end{equation}
which gives
\begin{equation}
  e^{\nu} = \frac{ c_0 \, e^{\lambda} \left[  e^{\lambda} - ( c_1 \, r^2 + 1) \right]}{ \left( \frac{4 c_2}{r^2} - 3 c_1 \right) e^{\lambda} - c_1 ( c_1 \, r^2 + 1) } \, .  \label{nde-nu}
\end{equation}
Thus, the physical variables $\rho, p$ and $\pi_{ab}$ for this case are
\begin{equation}
  \rho = \frac{ \left( \frac{4 c_2}{r^2} - 3 c_1 \right) - c_1 ( c_1 \, r^2 + 1) \, e^{-\lambda}  }{ e^{\lambda} -  c_1 \, r^2 - 1  }  \, , \qquad  p = \frac{1}{3} \left( c_1 \, e^{-\lambda} + \frac{2 c_2}{r^2} \right) \, ,
\end{equation}
\begin{equation}
  \pi_{11} = \frac{2}{3} \left( c_1 - \frac{c_2}{r^2} e^{\lambda} \right) \, , \qquad \pi_{22} = -\frac{r^2}{2} e^{-\lambda} \pi_{11}\, , \qquad  \pi_{33} = \sin^2 \theta \pi_{22} \, ,
\end{equation}
with the choice of timelike observers. Note that the prefect fluid which requires $\pi_{ab} = 0$  is not allowed in this case since Eqs. \eqref{nde-t0} - \eqref{nde-t2} are not equivalently satisfied for $e^{\lambda (r)} = c_1 \, r^2 / c_2$ and $\nu(r)$ that comes from \eqref{nde-nu}.

The vector fields for cases {\bf (ND-A)}-{\bf (ND-D)} are CMCs of the original metric \eqref{metric} which are \emph{fifteen} for $\alpha^2 \neq 0$ and \emph{twelve} for $\alpha^2 = 0$ in almost all cases. Also, we can employ an anisotropic fluid without heat flux for all cases in this section.

\section{Conclusions}
\label{conc}

Symmetries of the metric tensor on a manifold, like  KVs, HMs and CKVs have finite dimensional Lie algebras as the metric tensor is always non-degenerate. The maximum dimension for the Killing
algebras (in four dimensional space) is 10, for HMs it is 11 and for CKVs 15. But there is no such guarantee for other tensors which can be degenerate as well as non-degenerate. Thus the Lie algebras for RCs, MCs etc. can be finite as well as infinite. When the tensor is non-degenerate the algebra of RCs and MCs is finite and maximum dimension is 10, but for the degenerate tensor it can be finite as well as infinite. For conformal collineations the maximum dimension is 15 for the non-degenerate tensor. For the degenerate case finite dimensionality is not guaranteed.

In this work, we have completely classified CMCs for static spherically symmetric spacetimes which are not of Bertotti-Robinson type. We have seen in Section \ref{degenerate} that if the energy-momentum tensor is degenerate, i.e., $\det (T_{a b}) = 0$, then the CMCs have infinite degrees of freedom. In Section \ref{non-degenerate} in which the non-degenerate case where $\det (T_{a b}) \neq 0$ is considered, there are \emph{fifteen} CMCs if the separation constant is not zero, or \emph{twelve} CMCs if the separation constant vanishes. In the cases {\bf (ND-B-ii)} and {\bf (ND-E)} of \emph{six} dimensional Lie algebras the conformal factor comes out to be zero, and thus, they are actually MCs and not CMCs.

We point out that in degenerate case {\bf (D-A4-i)} we found an exact spherically symmetric solution of the form
\begin{eqnarray}
& & ds^2 =  \left[ -\frac{\nu_0}{4} \sqrt{1 - \frac{2 M }{r} } + \frac{\nu_1}{2} \left\{ r - 6 M + 3 M \sqrt{1 - \frac{2 M }{r} }  \ln \left( r + \sqrt{ r (r - 2 M)} - M \right) \right\} \right]^2  dt^2  \nonumber \\ & & \qquad \qquad - \frac{ dr^2}{1 - \frac{2 M}{r}} - r^2 \left( d\theta^2 + \sin^2 \theta \, d\phi^2 \right) \, ,
\end{eqnarray}
where we have taken $\lambda_1 = 2 M$. This new metric is {\it radiation-dominated fluid} solution for EFEs. It is interesting to note from Eq. \eqref{nec-anf} that the NEC for the above solution reads $p \geq 0$. For this metric the Ricci scalar becomes
\begin{equation}
  R = \frac{ 8 \nu_1}{ - \nu_0 \sqrt{r (r- 2 M)} + 2 \nu_1 \left[  r ( r - 6 M) + 3 M \sqrt{r (r- 2 M)} \ln \left( r + \sqrt{r (r- 2 M)} - M \right) \right] } \, .
\end{equation}
This metric reduces to the well-known form of Schwarzschild solution if $\nu_0 = -4$ and $\nu_1 = 0$. The above solution admits \emph{eleven} CMCs, which are given in \eqref{sphmin} and \eqref{a4i-cmc7}, and they have finite dimensional Lie algebra.

For the sake of completeness, we shortly touch on the possible extension of similar studies in generic dimensions. It is well-known that the spherically symmetric metrics continue to exist in space-times of dimension greater than four. However, it will be much more complicated to solve the symmetry equations even for static and spherically symmetric space-times in higher dimensions.


\section*{Acknowledgements}

KS acknowledges a research grant from the Higher Education Commission of Pakistan.


\end{document}